\documentclass[aps,prb,twocolumn,showpacs,amsmath,amssymb,superscriptaddress]{revtex4}
\usepackage{graphicx}
\usepackage{amssymb}
\usepackage{natbib}

\begin{document}
\title{Electron doping evolution of the neutron spin resonance in NaFe$_{1-x}$Co$_{x}$As}

\author{Chenglin Zhang}
\affiliation{Department of Physics and Astronomy, Rice University, Houston, Texas 77005, USA}

\author{Weicheng Lv}
\affiliation{Department of Physics and Astronomy, Rice University, Houston, Texas 77005, USA}

\author{Guotai Tan}
\affiliation{Department of Physics, Beijing Normal University, Beijing 100875, China}

\author{Yu Song}
\affiliation{Department of Physics and Astronomy, Rice University, Houston, Texas 77005, USA}

\author{Scott V. Carr}
\affiliation{Department of Physics and Astronomy, Rice University, Houston, Texas 77005, USA}

\author{Songxue Chi}
\affiliation{Quantum Condensed Matter Division, Oak Ridge National Laboratory, Oak Ridge, Tennessee 37831, USA}

\author{M. Matsuda}
\affiliation{Quantum Condensed Matter Division, Oak Ridge National Laboratory, Oak Ridge, Tennessee 37831, USA}

\author{A. D. Christianson}
\affiliation{Quantum Condensed Matter Division, Oak Ridge National Laboratory, Oak Ridge, Tennessee 37831, USA}

\author{ J. A. Fernandez-Baca}
\affiliation{Quantum Condensed Matter Division, Oak Ridge National Laboratory, Oak Ridge, Tennessee 37831, USA}
\affiliation{Department of Physics and Astronomy, The University of Tennessee, Knoxville, Tennessee 37996, USA}

\author{L. W. Harriger}
\affiliation{NIST Center for Neutron Research, National Institute of Standards and Technology, Gaithersburg, Maryland 20899, USA}

\author{Pengcheng Dai}
\email{pdai@rice.edu}
\affiliation{Department of Physics and Astronomy, Rice University, Houston, Texas 77005, USA}
\affiliation{Department of Physics, Beijing Normal University, Beijing 100875, China}

\begin{abstract}
Neutron spin resonance, a collective magnetic excitation coupled to superconductivity, is one of the most prominent features shared by a broad family of unconventional superconductors including copper oxides, iron pnictides, and heavy fermions. 
In this work, we study the doping evolution of the resonances in NaFe$_{1-x}$Co$_x$As covering the entire superconducting dome. For the underdoped compositions, two resonance modes coexist. As doping increases, the low-energy resonance gradually loses its spectral weight to the high-energy one but remains at the same energy. By contrast, in the overdoped regime we only find one single resonance, which acquires a broader width in both energy and momentum, but retains approximately the same peak position even when $T_c$ drops by nearly a half 
compared to optimal doping. These results suggest that the energy of the resonance in electron overdoped NaFe$_{1-x}$Co$_x$As 
is neither simply proportional
to $T_c$ nor the superconducting gap, but is controlled by the multi-orbital character of the system 
and doped impurity scattering effect.
\end{abstract}


\maketitle


\section{introduction}
Although the microscopic origin of superconductivity remains unresolved nearly 30 years after the discovery of 
high-transition temperature 
(high-$T_c$) copper oxides \cite{keimer}, 
it is generally believed that 
spin fluctuation mediated electron pairing is a common thread for unconventional superconductors including copper oxide, iron-based, and heavy-fermion superconductors \cite{scalapino,pdai}. Regardless of the dramatic differences 
in the ground states of their parent compounds and the microscopic origins of magnetism in different families of unconventional
superconductors, 
inelastic neutron scattering experiments have revealed that superconductivity induces a collective magnetic excitation, 
termed neutron spin resonance, near the antiferromagnetic (AF) ordering wave vector of their parent compounds 
\cite{mignod,Eschrig,jmtranquada,dai,christianson}.
Experimentally, the resonance occurs at an energy $E_r$ and enhances dramatically below $T_c$ like the superconducting order parameter. 
In the Fermi surface nesting (itinerant electron) picture \cite{Eschrig}, the resonance is a spin-exciton mode in the particle-hole channel.
If the superconducting order parameter has a sign-change below $T_c$,
 the dynamic spin susceptibility will develop a pole, namely the resonance, at an energy $E_r$ below the particle-hole continuum $2\Delta$ (where $\Delta$ is the superconducting gap) \cite{Eschrig}.  In the case of iron pnictide 
superconductor NaFe$_{1-x}$Co$_x$As with hole and electron
Fermi surfaces at $\Gamma$ and $M$ points, respectively [Fig. 1(a),1(e)-1(g)] \cite{stewart,hirschfeld,slli09,Liu_arpes,thirupathaiah,GTTan2013,qqge}, the resonance arises from quasiparticle excitations between the 
sign-reversed hole and electron Fermi surfaces and occurs at an energy below the sum of their superconducting 
gap energies ($E_r\le \Delta_h+\Delta_e=2\Delta$, where $\Delta_h$ and $\Delta_e$ are superconducting gaps 
at hole and electron Fermi surfaces, respectively) [Fig. 1(b)-1(d)] \cite{Korshunov,Maier}.
Although $T_c$ differs dramatically for copper oxide, iron-based, 
and heavy-fermion superconductors, the resonance energy $E_r$ is approximately related to $T_c$ via  
 $E_r/k_BT_c \approx 4-6$ or the superconducting gap energy $\Delta$ via $E_r/2\Delta = 0.64$ ($2\Delta/k_B T_c = A + B T_c$, where $A$ and $B$ are constants.) \cite{miaoyin2010,Inosov2011,GYu}. 
While these results suggest that the resonance may be a common thread for unconventional superconductors \cite{scalapino}, 
most of the inelastic neutron scattering measurements on the resonance in iron pnictides 
are focused on underdoped and optimally doped samples with few 
experiments on overdoped regime of the phase diagram  \cite{Inosov2011}.

\begin{figure}[t] 
\includegraphics[scale=.38]{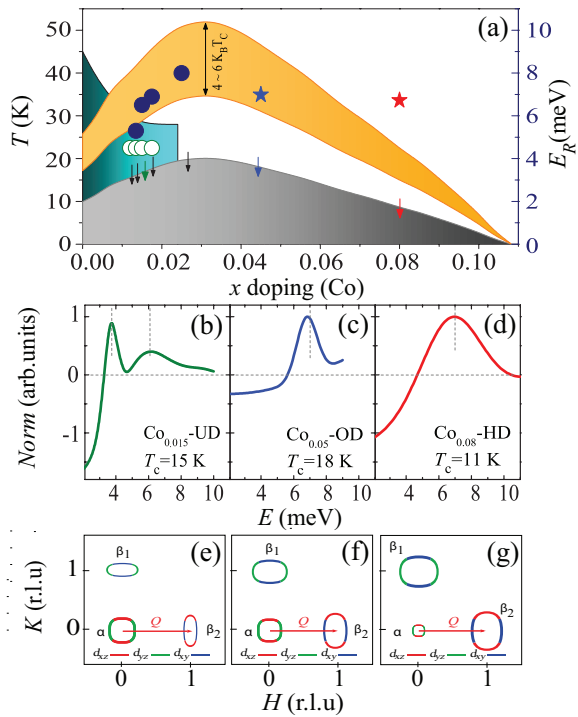}
\caption{(Color online)
(a) The electronic phase diagram of NaFe$_{1-x}$Co$_x$As, where the arrow indicates the Co-doping levels studied 
in this work.  The grey shaded area marks the Co-doping dependence of $T_c$. The 
region with AF order is represented by the green shaded area.  The open circles are energies of the first resonance $E_{r1}$, and 
the filled circles and stars are energies of the second resonance $E_{r2}$. The yellow shaded area
indicates approximate range of $E_r\approx 4-6 k_BT_c$ obtained from previous work \cite{miaoyin2010,Inosov2011}.
(b-d) The schematic energy dependence of the resonance for three characteristic Co-doping levels, including
underdoped (UD), optimally doped (OD), and highly overdoped (HD).
(e-g) Schematic plots of the Fermi surfaces for the above three compositions. 
The color indicate different orbitals. The anisotropic superconducting gap $\Delta_e$ on the electron pockets 
in the underdoped compounds become isotropic on the overdoped side \cite{qqge}.
}
\end{figure}

In this paper, we report systematic inelastic neutron scattering studies of 
the resonance in iron pnictide superconductors NaFe$_{1-x}$Co$_{x}$As 
for Co compositions throughout the entire superconducting dome \cite{Parker10,Wright,AFWang12,AFWang13,GTTan2016}. In previous work on electron 
underdoped NaFe$_{0.985}$Co$_{0.015}$As where static AF order coexists with superconductivity ($T_N=30$ K, and $T_c=15$ K), we find a 
dispersive sharp resonance near $E_{r1}=3.25$ meV and a broad dispersion-less
mode at $E_{r2}=6$ meV at the AF ordering 
wave vector ${\bf Q}_{\rm AF}$ \cite{slli09,CLZhang13}. Upon moving to electron overdoped NaFe$_{0.955}$Co$_{0.045}$As
 without static AF order ($T_c=20$ K), there is only one sharp resonance at $E_r=7$ meV \cite{clzhang13b}.
By carrying out systematic measurements on NaFe$_{1-x}$Co$_{x}$As with nominal Co-doping of 
$x=0.012,0.0135,0.0175,0.025,0.08$ [Fig. 1(a)], we establish 
the electron-doping evolution of the resonance throughout the superconducting phase.  In the underdoped regime, we confirm the earlier results showing the presence of double resonance peaks at $E_{r1}$ and $E_{r2}$ as shown in Fig. 1(b). 
 As doping increases, $E_{r1}$ stays almost the same value while $E_{r2}$ moves to higher energies. 
At optimal doping and in slightly overdoped samples, the low-energy resonance disappears
and only a single sharp resonance occurs at $E_{r2} = 7$ meV [Fig.~1(c)]. 
For heavily overdoped $x=0.08$, the resonance 
becomes much broader in energy but retains its peak position [Fig.~1(d)]. 
These results indicate that the resonance energy in the electron 
overdoped regime is neither directly associated with $T_c$ via the empirical relation 
$E_r / k_B T_c = 4 \sim 6$ nor with $\Delta$ via $E_r/(\Delta_h+\Delta_e) = 0.64$ \cite{miaoyin2010,Inosov2011,GYu},
thus suggesting that the multi-orbital character and the inter-band nonmagnetic 
impurity scattering due to Co-doping in NaFe$_{1-x}$Co$_{x}$As play an important role in 
determining the properties of the resonance. 

\begin{figure}[t] 
\includegraphics[scale=.38]{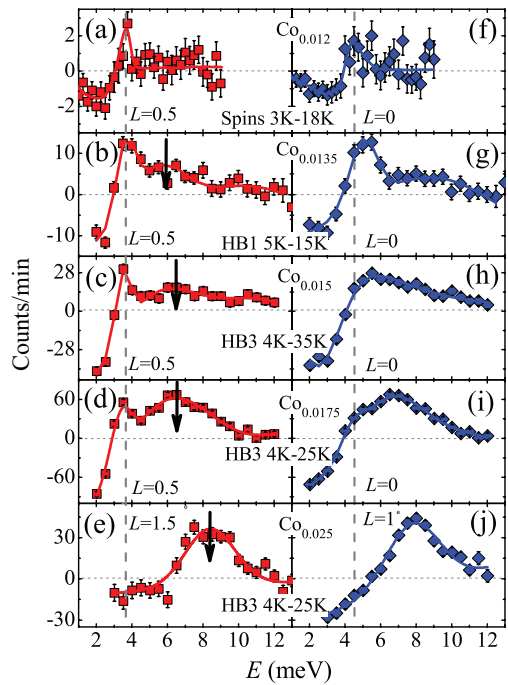}
\caption{(Color online)
The neutron resonances in NaFe$_{1-x}$Co$_x$As as a function of increasing $x$, 
obtained as the difference of the energy-scans above and below $T_c$ at 
the wave vectors ${\bf Q}_{\rm AF}=(1,0,L)$ with $L=0.5,1.5$ (a-e) and $\bm{Q}=(1,0,L)$ with
$L=0,1$ (f-j). (a,f) $x = 0.012$ (UD); (b,g) $x = 0.0135$ (UD); 
(c,h) $x = 0.015$ (UD); (d,i) $x = 0.0175$ (UD); (e,j) $x = 0.025$ (OP). 
The plots are obtained directly by subtracting the superconducting state energy scan from those in the normal state without correcting
for background, as is commonly done for determining the energy of the resonance \cite{pdai,mignod,Eschrig,jmtranquada}. 
The solid lines are fits with two Gaussians. 
The vertical dashed lines denote the low-energy resonance at $E_{r1}$ in (a-d) and (f-i).
The negative intensity below the resonance indicates the opening of a spin gap below $T_c$.  
The vertical arrows indicate the peak positions of the high-energy resonance $E_{r2}$ at ${\bf Q}_{\rm AF}=(1,0,L)$
with $L=0.5,1.5$.
}
\end{figure}

\section{Experimental results}

We grew single crystals of NaFe$_{1-x}$Co$_x$As by self-flux method as described before \cite{Spyrison}.
The sample quality has been characterized by various techniques, which found that 
bulk superconductivity appears in the doping range of $ 0.012 \leq x \leq 0.1$ \cite{GTTan2013}. 
Our inelastic neutron scattering experiments were carried out over the entire doping range as shown by vertical arrows in 
Fig. 1(a). The measurements were performed on the HB-1 and HB-3 
thermal triple-axis spectrometers at High Flux Isotope Reactor, Oak Ridge National Laboratory, and SPINS cold triple-axis spectrometer at the NIST Center for Neutron Research.
Pyrolytic graphite (PG) monochromator and analyzer were used with fixed final neutron energies at 
$E_f=14.7$ meV and $E_f=5$ meV for thermal and cold neutron measurements, respectively.
The corresponding energy resolutions are $\Delta E \approx 1.2$ meV and 
$\Delta E \approx 0.15$ meV, respectively, at the AF ordering elastic position.
Several pieces of crystals co-aligned with a total mass of $\sim$ 10 g and 
the mosaic of $\sim 3^\circ$ were used in each experiment.  
The wave vector $\bm{Q}$ at $(q_x, q_y, q_z)$ in \AA$^{-1}$ is defined as 
$(H, K, L) = (q_xa/2\pi, q_yb/2\pi, q_zc/2\pi)$ in reciprocal lattice unit (r.l.u) using the orthorhombic unit cell
where $a\approx b\approx 5.589$ \AA\ and $c \approx 6.980$ \AA\ at 3 K. The samples are aligned in the $[H,0,L]$ scattering zone, 
where the resonance occurs at the AF wave vector ${\bf Q}=(1,0)$, consistent with the Fermi surface
nesting wave vector shown in Fig. 1(e) \cite{CLZhang13,clzhang13b}. Some measurements are carried out in 
the $[H,K,0]$ scattering plane.

\begin{figure}[t] 
\includegraphics[scale=.38]{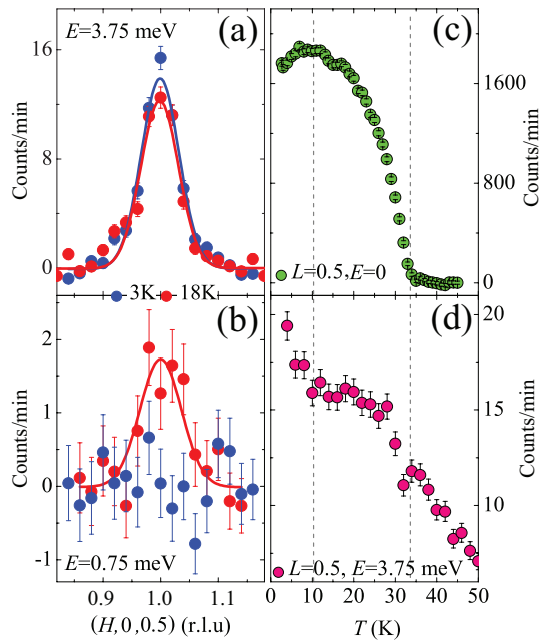}
\caption{(Color online) Wave vector and temperature dependence of the magnetic scattering for 
NaFe$_{1-x}$Co$_x$As with $x=0.012$ when the system is near bulk superconductivity. 
The sample has $T_N$ of $\sim$35 K and $T_c \approx 11$ K. The data was collected on SPINS.
}
\end{figure}

To systematically investigate the electron-doping evolution of the double resonance in the underdoped regime \cite{CLZhang13}, we first 
focus on a series of 
compositions from $x = 0.012$ to $x = 0.0175$ [Fig.~2(a)-(d),(f)-(i)]. Similar to previous neutron scattering 
work \cite{mignod,miaoyin2010}, we define resonance as the intensity gain of magnetic scattering in the superconducting state.  For this purpose, energy scans
are carried out at fixed wave vectors below and above $T_c$, and the net intensity gain of the scattering below $T_c$ is ascribed to the resonance.
In the case of NaFe$_{1-x}$Co$_{x}$As, previous work has shown that the resonance occurs at slightly different energies 
at the AF zone center ${\bf Q}_{\rm AF}=(1,0,L)$ with $L=0.5,1.5$ and zone boundary with $L=0,1$ \cite{CLZhang13}.  We have therefore carried out systematic measurements at these two wave vectors for all Co-doping levels.  Figure 2(a) and 2(f) shows the outcome
for NaFe$_{1-x}$Co$_{x}$As with $x = 0.012$, when the system first becomes near the bulk superconducting phase \cite{GTTan2013}. 
The temperature difference plot shows a resonance peak at $E_{r1}=3.75$ meV for $L=0.5$ and $E_{r1}=4.5$ meV for $L=0$ with the corresponding 
spin gaps of $E_g\approx 3$ and 4 meV, respectively. To further confirm the existence of the resonance, we carried out
momentum and temperature dependence measurements on SPINS. Figure 3(a) shows constant-energy scans at  
$E_{r1}=3.75$ meV along the $[H,0,0.5]$ direction, which reveals clear intensity gain below $T_c$ at 
${\bf Q}_{\rm AF}$.  For an energy below the resonance at $E=0.75$ meV, AF spin fluctuations are completely
suppressed below $T_c$, suggesting the opening of a spin gap in the superconducting state [Fig. 3(b)].
Temperature dependence of the elastic magnetic scattering is shown in Fig. 3(c).  Similar to previous work 
on underdoped superconducting iron pnictides \cite{Pratt09,Christianson09}, we see clear evidence for AF order below
$T_N\approx 35$ K and the suppressive effect of superconductivity on AF order.  Figure 3(d) shows temperature dependence of scattering
at $E_{r1}=3.75$ meV and  ${\bf Q}_{\rm AF}=(1,0,0.5)$.  Based on these results, we find 
clear evidence for the resonance in the $x=0.012$ compound.
 
\begin{figure}[t] 
\includegraphics[scale=.45]{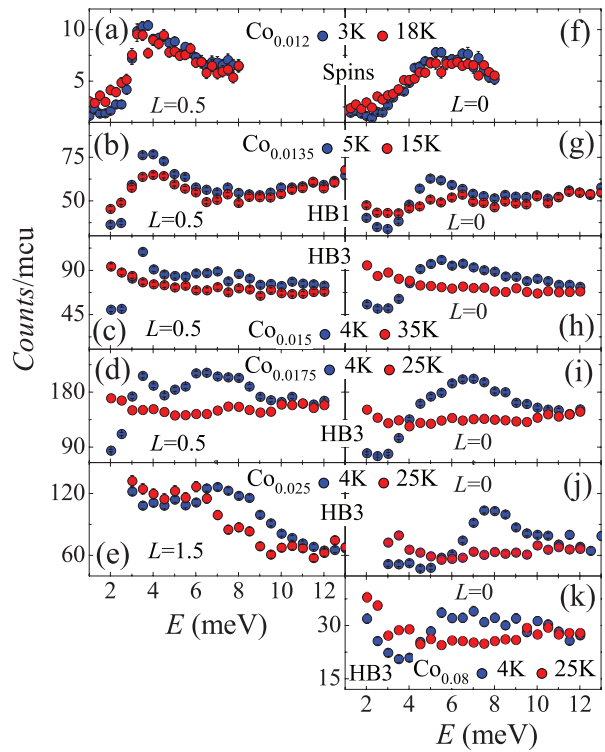}
\caption{(Color online) The raw data of energy scans at wave vectors ${\bf Q}_{\rm AF} = (1,0,L)$ 
with $L= 0.5,1.5$ (a-e) and $L= 0$ (f-k) obtained for NaFe$_{1-x}$Co$_x$As with different $x$ 
above and below $T_c$. }
\end{figure}

At higher doping levels, $x = 0.0135$ [Fig.~2(b, g)], $x = 0.015$ [Fig.~2(c, h)], and $x = 0.175$ [Fig.~2(d, i)], a second resonance mode with a broad width appears at a higher energy $E_{r2}$. 
As the superconducting transition temperature $T_c$ increases with increasing Co-doping, $E_{r2}$ also increases, whereas $E_{r1}$ stays at almost the same energy for ${\bf Q}_{\rm AF}=(1,0,0.5)$. These results suggest that the energy of the first resonance is not directly associated with $k_BT_c$. Furthermore, we note that the spectral weight of the low-energy resonance 
gradually shifts to the high-energy one with increasing Co-doping. 
Near optimal doping $x = 0.025$ ($T_c = 22$ K) [Fig.~2(e, j)] \cite{GTTan2013}, the low-energy resonance completely vanishes and only the high-energy resonance is present. 
Comparing the left and right panels of Fig.~2, we see that in the underdoped regime, the energy of the first resonance 
shows similar out-of-plane momentum dependence as in the underdoped superconducting BaFe$_2$As$_2$ systems doped with
Co, Ni, and P \cite{lee2013universality}, being higher at $L = 0$ than at $L = 0.5$. 
Near optimal superconductivity, the resonance energy becomes dispersion-less, occurring at the same energy for  
both $L=0.5$ and 1. Figure 4 shows the raw data below and above $T_c$ for different Co-doping 
samples obtained at various triple-axis spectrometers. Although energy dependence of the spin excitations spectra are some what different
in the underdoped samples where superconductivity coexists with static AF order and optimally/overdoped samples where there are no
static magnetism, we collected the data below and above $T_c$ to determine accurately the effect of superconductivity
on the magnetic excitations spectra.

\begin{figure}[t] 
\includegraphics[scale=.35]{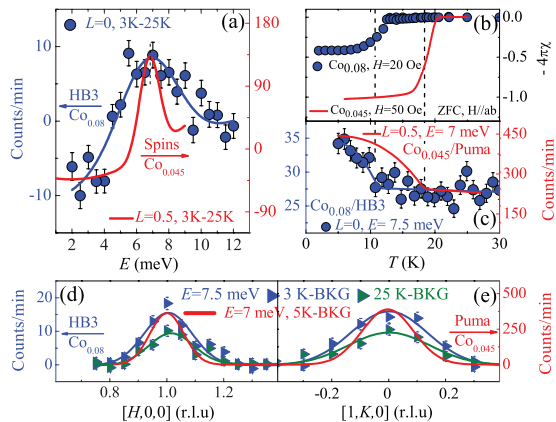}
\caption{(Color online)
Comparison of the resonance in the overdoped regime at two compositions $x = 0.045$ and $x = 0.08$. 
(a) The difference of the energy scans above and below $T_c$ normalized by the corresponding peak intensities. The $x = 0.045$ and $x = 0.08$ compositions have similar peak energies around 7 meV, but have very different energy widths. For Co-doping levels above $x = 0.025$, resonances are not dispersive along the $L$ direction. (b) Temperature dependence of the susceptibility for $x = 0.045$ and $x = 0.08$.
The superconducting volume fraction of the $x = 0.08$ sample is about 40\%.
(c) Temperature dependence of the scattering intensity at the resonance energy, which show the onset of the resonance modes at their respective transition temperatures $T_c$. (d,e) The wave vector scans at the resonance energies along the 
$[H , 0, 0]$ and $[1, K, 0]$ directions below and above $T_c$ for $x=0.08$. 
Similar data for $x=0.045$ at 5 K is shown in red solid line \cite{clzhang13b}.
The blue and green solid lines are Gaussian fits to the data.
}
\end{figure}

Figure 5 summarizes the results for an electron-overdoped sample with $x=0.08$ ($T_c=11$ K).
Since $T_c$ of the sample is significantly lower than that of the electron doped $x=0.045$
[Fig. 5(b)] \cite{clzhang13b}, we would expect a reduction in the 
superconducting gap amplitude $2\Delta=\Delta_h+\Delta_e$ as well \cite{XDZhou}. If the resonance is a bound-state below the 
particle-hole continuum $2\Delta$ \cite{Eschrig}, 
there should be a corresponding reduction in the mode energy on moving from $x=0.045$ to $x=0.08$.
Figure 5(a) compares temperature difference plot of the energy scans below and above $T_c$ for $x=0.045$ to $x=0.08$.
While there is a clear resonance in both samples, the resonance for $x = 0.08$ shows a much broader width compared to that of $x = 0.045$ even
considering the differences in instrumental energy resolution in these two experiments.
In addition, the two resonances have almost the same peak energy at $E_r = 7$ meV, despite the large reduction in $T_c$ from $x=0.045$ to $x=0.08$. 
To confirm that the intensity gain below $T_c$ in the $x=0.08$ sample is indeed the resonance, we show in Fig. 5(c) 
temperature dependence of the scattering at $E_r = 7$ meV. For both $x=0.045$ to $x=0.08$ samples, there are 
clear superconducting order parameter like intensity gain below $T_c$'s, a hallmark of the resonance.
Figure 5(d) and 5(e) shows constant-energy scans above background below and above $T_c$ 
along the $[H,0,0]$ and $[1,K,0]$ directions, respectively, for $x=0.08$. The red solid lines are similar wave vector scans for 
the $x=0.045$ sample \cite{clzhang13b}.
These results confirm the temperature difference plots, showing that intensity gain of 
below $T_c$ in Fig. 5(a) and 5(c) is indeed from the resonance.  Although $x=0.08$ sample is not a 100\% bulk 
superconductor [Fig. 5(b)], the differences
between the superconducting and normal state should still represent the effect of superconductivity to the magnetic excitations. 
Based on the properties of the resonance 
in the $x=0.045$ to $x=0.08$ samples shown in Fig. 5, we conclude that the mode energy 
 $E_r$ does not scale linearly with $T_c$ or $\Delta$. 
The ratios $E_r/ k_B T_c$ and $E_r/ 2 \Delta$ in the $x=0.08$ composition are well above the values proposed in the universal relations [see Fig.~1(a)]. Furthermore, we find that while the resonance for both samples are centered at the AF ordering wave vector, 
the $x= 0.08$ sample has considerable broader ${\bf Q}$-width along the $H$ and $K$ directions. 

\begin{figure}[t] 
\includegraphics[scale=.55]{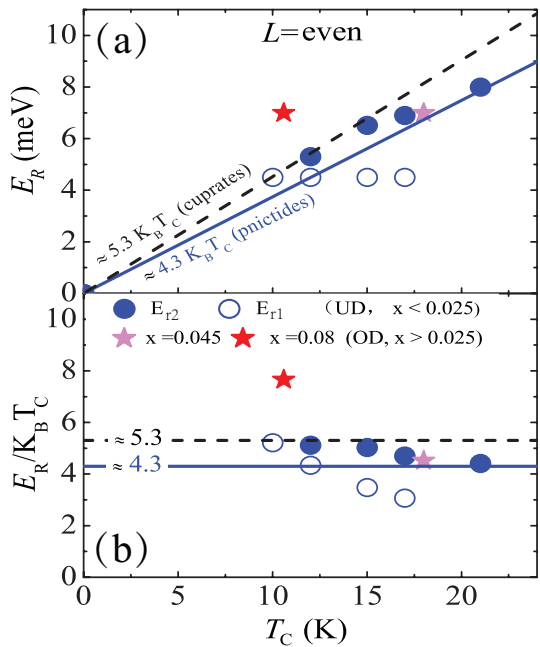}
\caption{(Color online)
(a) The Co-doping dependence of the resonance energy $E_{r1}$ and $E_{r2}$ as a function of 
$T_c$ in NaFe$_{1-x}$Co$_x$As.  The solid and dashed lines are expected values for copper oxide and iron pnictide
superconductors \cite{Inosov2011,GYu}.
(b) The ratio $E_r/k_BT_c$ as a function of $T_c$. The lines represent the expected linear relations. The low-energy resonance $E_{r1}$ in the underdoped regime and the high-energy resonance $E_{r2}$ in the overdoped composition do not follow the expected behavior.
}
\end{figure}

\section{Discussion and Conclusion}

Figure 6 summarizes the Co-doping evolution of the resonance in NaFe$_{1-x}$Co$_{x}$As.  The open circles in
Fig. 6(a) shows that the energy of the first resonance $E_{r1}$ is essentially 
 $T_c$ independent.  If the double resonance originates from the superconducting gap anisotropy in the
underdoped regime \cite{qqge,CLZhang13,RYu14}, one would 
expect that $E_{r1}$ decreases with increasing doping, contrary to the observation. 
On the other hand, these results may indicate 
that the first resonance is coupled with the static AF order and spin waves as suggested theoretically \cite{WCLv14}.
If this is indeed the case, one would expect that an uniaxial pressure used to detwin the sample would separate the double resonance, where the first resonance associated with spin waves ($E_{r1}$) should appear at ${\bf Q}_{\rm AF}=(\pm 1,0)$ but not at $(0,\pm 1)$, while the second 
resonance ($E_{r2}$) arising from Fermi surface nesting and itinerant electron would appear at both  
${\bf Q}_{\rm AF}=(\pm 1,0)$ and $(0,\pm 1)$ wave vectors \cite{clzhang2015}. 
However, our recent neutron scattering experiments on uniaxial pressure detwinned sample found double resonance at both wave vectors, thus suggesting that the first mode cannot be associated with spin waves at ${\bf Q}_{\rm AF}=(\pm 1,0)$ \cite{clzhang2015}.
While these results seem to rule out the AF order origin for the first resonance, 
a more detailed investigation using superconducting gap anisotropy scenario is necessary to 
determine if such a model can explain our observation \cite{RYu14}. 
The solid circles and stars in Fig. 6(a) show the $T_c$ dependence of the second 
resonance energy $E_{r2}$. While the mode energies for underdoped and slightly overdoped samples fall
within the generally accepted values of $E_r\propto k_BT_c$, the resonance energy for $x=0.08$ clearly deviates from
the expectation.  Figure 6(b) plots the same data in terms of $E_r/k_BT_c$.

To understand the behavior of the resonance in the electron-overdoped regime of NaFe$_{1-x}$Co$_{x}$As, 
we consider two essential effects from Co-doping. The first one is the introduction of additional electron charge carriers, which causes the hole pockets to shrink and the electron pockets to expand, as illustrated in Fig.~1(e-g). 
As the mismatch between the electron and hole pockets increases with doping, the resonance peak obtains more contributions from the scattering momenta that are away from the AF order wave vector $(1,0)$, and therefore shows a broader peak in the momentum space. This is reminiscent of 
of the wave vector dependence of the resonance in BaFe$_{2-x}$Ni$_x$As$_2$ family of materials, where the mode becomes
transversely incommensurate in the electron-overdoped regime \cite{HQLuo}, except here the scattering is
commensurate in the entire measured doping range. With electron overdoping and sinking of the hole pocket below Fermi surface,
the low-energy spin excitations vanish together with the suppression of superconductivity \cite{Carr2016}, very similar to the
presence of a large spin gap in electron-overdoped nonsuperconducting BaFe$_{1.7}$Ni$_{0.3}$As$_2$ \cite{mwang13}.
The second less considered effect is that the Co dopants can also act as local nonmagnetic impurities. 
In iron pnictides where the superconducting order parameter changes sign between the hole and electron pockets [Fig. 1(e-g)], interband scatterings from these impurities are superconducting pair-breaking.
Therefore, as more impurities are introduced with increasing Co-doping, we expect that the superconducting
gap to be gradually filled and the critical temperature $T_c$ to be reduced due to these pair-breaking scatterings. However, the spin resonance arises from the superconducting quasiparticles that retain the original gap amplitude $\Delta$.
Therefore, the resonance energy $E_r$ is not much affected by these interband nonmagnetic scatterings, and the mode
will acquire a larger width in energy due to the broadened quasiparticle peak with increasing impurity concentration \cite{Maiti}.
 These results are 
consistent with our experimental observations, suggesting the important roles of the impurity scatterings
in determining the energy and wave vector dependence of the resonance. Our study in the overdoped NaFe$_{1-x}$Co$_x$As have demonstrated that the Co dopants introduce two important effects into the system, namely the additional itinerant electrons and local nonmagnetic impurities.

\section{Acknowledgments}
We thank Caleb Redding and Z. C. Sims for their help in single crystal growth efforts.
The single crystal growth and neutron scattering
work at Rice is supported by the U.S. DOE, BES under contract no. DE-SC0012311 (P.D.).
Part of the work is also supported by the Robert A. Welch foundation grant numbers C-1893 (P.D.).
The use of Oak Ridge National Laboratory’s High Flux Isotope Reactor was sponsored by the Scientific User Facilities Division, Office of Basic Energy Sciences, U.S. DOE.

\end{document}